\documentclass[12pt,showpacs,preprintnumbers,nofootinbib,preprint]{revtex4}%
\usepackage{amsmath}
\usepackage{graphicx}
\usepackage{bm}
\usepackage{amsfonts}
\usepackage{amssymb}%
\begin{document}
\preprint{ }
\title{About a peculiar extra $U(1):Z^{\prime}$ discovery limit\\Muon anomalous magnetic moment \\Electron electric dipole moment }
\author{Jae Ho Heo}
\email{jheo1@uic.edu}
\affiliation{Physics Department, University of Illinois at Chicago, Chicago, Illinois
60607, USA }

\begin{abstract}
The model (Lagrangian) with a peculiar extra $U(1)$ \cite{Bar05} is clearly
presented. The assigned extra $U(1)$ gauge charges give a strong constraint to
build Lagrangians. The $Z^{\prime}$ discovery limits are estimated and
predicted at the Tevatron and the LHC. The new contributions of the muon
anomalous magnetic moment are investigated at one and two loops, and we
predict that the deviation from the standard model may be explained. The
electron electric dipole moment could also be generated because of the
explicit CP violation effect in the Higgs sector, and a sizable contribution
is expected for a moderately sized CP phase (argument of the CP-odd Higgs),
$0.1\leq\sin\delta\leq1$ $(6^{\circ}\leq\arg(A)\leq90^{\circ})$.

\end{abstract}

\maketitle

\section{Introduction}

An extra $U(1)$ (or a few extra $U(1)^{\prime}$s) may arise in the context of
grand unified theories \cite{Agu90}, superstring theories \cite{Ros88} or
generically emerge as simple extensions of the standard model (SM). Therefore,
the models with an extra $U(1)$ (or a few extra $U(1)^{\prime}$s) have been
extensively considered. Recently, Barr and Dorsner \cite{Bar05} suggested
another possibility for an extra $U(1)$ gauge, which satisfies all anomaly
constraints in a maximally economical way, whatever its origin\footnote{The
origin is close to the Pati-Salam model with an extra $U(1)$\cite{Bar05}.} is.
In the standard model, all the possible anomalies from triangle diagrams of
three gauge bosons must be canceled if the Ward identities of the gauge theory
are to be satisfied. The existence of an extra $U(1)$ brings six additional
anomaly cancellation conditions, $U(1)_{Y}^{2}\times U(1)_{X}$, $U(1)_{Y}%
\times U(1)_{X}^{2}$, $U(1)_{X}^{3}$, $SU(2)^{2}\times U(1)_{X}$,
$SU(3)^{2}\times U(1)_{X}$ , gravity $\times U(1)_{X}$. These anomaly
cancellations are nontrivial\footnote{The general analyses about these
cancellations can be found in Ref.\cite{Ema02}.}, but Barr and Dorsner showed
a remarkably trivial solution \cite{Bar05} with a single extra lepton triplet
per family. These gauge anomalies are exactly canceled for the fermion gauge
charges listed at Table I.

In this letter the model (Lagrangian) is clearly presented. With an extra
lepton triplet, an additional Higgs singlet is necessary to provide masses of
the exotic leptons and the extra gauge boson $Z^{\prime}$. The Higgs singlet
would involve the extra $U(1)$ gauge symmetry breaking, and we assume that the
symmetry is broken near the weak scale. A Higgs triplet\footnote{The Higgs
triplet is introduced because of the need to induce the CP violating
interaction in this work. We can add additional scalars, such as nongauged
Higgs singlets or doublets, in other ways, but adding the gauged scalars is
more generic. If we only consider neutrino mass generation without electric
dipole and dark matter (we need to impose the discrete $Z_{2}$ symmetry ; see
the next section) phenomenology, the additional Higgs triplet is unnecessary.
The interactions induced by the Higgs triplet could also give a significant
contribution to the anomalous magnetic moment of the muon (see Sec. IV).} with
the required gauge charges is added to explain our interesting phenomenology.
The new gauge boson $Z^{\prime}$ that generically emerges as gauging an extra
$U(1)$ is the intrinsic particle that explains the existence of an extra
$U(1)$, so its discovery limits at the Tevatron and LHC are estimated and
predicted. The muon anomalous magnetic moment, $a_{\mu}\equiv(g_{\mu}-2)/2$,
has been a powerful tool to account for new physics because of its importance.
We investigate the contributions involving the new particles at the one- and
two-loop levels.$\ $One can also see the explicit CP violation that generates
the electric dipole moment (EDM) of the electron $d_{e}$, and a sizable
contribution is expected via Barr-Zee two-loop mechanism for the moderate size
of the CP-phase (argument of the CP-odd Higgs), $0.1\leq\sin\delta\leq1$
$(6^{\circ}\leq\arg(A)\leq90^{\circ})$.

\begin{table}[t]
\caption{ Fermion gauge charges. $T_{3}$ is the weak isospin, $Y$ is the
hypercharge, $X$ is the extra $U(1)_{X}$ charge, and $Q=T_{3}+Y$ is the
electric charge. The charges for the right handed fermions can also be
assigned \ in the identical way. ($f_{L}^{c}\equiv(f_{L})^{c}$ in this letter,
so $(f^{c})_{L}$ implies the antiparticle of $f_{R}$).}%
\label{Table1}%
\begin{ruledtabular}
\begin{tabular}{ccccccccccc}
& $u_L$ &$d_L$ &$(u^c)_L$ &$(d^c)_L$ & $\nu_L$ &${\ell}_L$ &$({\ell}^c)_L$ & $E^+_L$ &$E^0_L$ &$E^-_L $\\
\hline
{\it T}$_3$ & ${1 \over 2}$  & $-{1 \over 2}$ & 0 & 0 & ${1 \over 2}$ & $-{1 \over 2}$& 0 & 1 & 0 & $-1$ \\
{\it Y} &${1 \over 6}$ & ${1 \over 6}$ & $-{2 \over 3}$ & ${1 \over 3}$ & $-{1 \over 2}$ & $-{1 \over 2}$& 1 & 0 & 0 & 0 \\
{\it X} & 1& 1 &$-1$  &$-1$  &1  & 1 & 1 &$ -1$ & $-1$ &$ -1$ \\
\end{tabular}
\end{ruledtabular}\end{table}

\section{Model description (Lagrangian)}

With the new particle content, the Yukawa potential for the lepton sector can
have the following enlarged form without the \textit{ad hoc} imposition of
lepton number conservation.%

\begin{equation}
y_{1}\mathrm{{Tr}}(\overline{E_{L}^{c}}E_{L})\eta+y_{2}\overline{L_{L}}%
\phi\ell_{R}+y_{3}\overline{L_{L}^{c}}i\sigma_{2}\chi L_{L}+y_{4}%
\overline{L_{L}^{c}}i\sigma_{2}E_{L}\phi+y_{5}\mathrm{{Tr}}(\overline{E_{L}%
}\chi)\ell_{R}+h.c.,
\end{equation}
where $\eta$ and $\chi$ denote a Higgs singlet and a Higgs triplet,
$\phi=(\phi^{+},\phi^{0})^{T}$ is a Higgs doublet, and $L=\left(  \nu_{\ell
},\ell\right)  ^{T}$ is the lepton doublet. The bi doublet representation is
taken for the additional lepton triplet and the Higgs triplet is also taken in
the form of a $2\times2$ matrix transforming under $SU(2)$ as $\chi\rightarrow
U\chi U^{\dag}$.%

\begin{equation}
E_{L}=%
\begin{pmatrix}
\frac{1}{\sqrt{2}}E^{0} & E^{+}\\
E^{-} & -\frac{1}{\sqrt{2}}E^{0}%
\end{pmatrix}
_{L},\text{ \ }\chi=%
\begin{pmatrix}
\frac{1}{\sqrt{2}}\chi^{+} & \chi^{++}\\
\chi^{0} & -\frac{1}{\sqrt{2}}\chi^{+}%
\end{pmatrix}
.
\end{equation}
The lepton triplet must be a Majorana combination. It should be noted that the
antisymmetric tensor $i\sigma_{2}$ follows from the antisymmetric property of
the charge conjugation.

Since the gauge charges of the leptons are already assigned by anomaly
constraints, the gauge charges of the Higgses are assigned by the combinations
with leptons in the Yukawa potential under the $SU(2)_{L}\times U(1)_{Y}\times
U(1)_{X}$ \ gauge invariance. If we introduce the assignment of $U(1)_{X}$
charges for the Higgses, the singlet $\eta$ must have a $U(1)_{X}$ charge of
$2$ from the $y_{1}$-term since $E$ has $-1$; the doublet $\phi$ may have a
charge of $2$ from the $y_{2}$-term and $0$ from the $y_{4}$-term ; and the
triplet $\chi$ may have a charge of $-2$ from the $y_{3}$-term and $0$ from
the $y_{5}$-term. The other gauge charges may be assigned in an analogous way,
and the assigned charges of the Higgses are listed at Table II. Notice that
the Higgs doublet and triplet may have two distinctive $U(1)_{X}$ charges.

\begin{table}[t]
\caption{ Higgs gauge charges. $T_{3}$ is the weak isospin, Y is the
hypercharge, $X$ is the $U(1)_{X}$ charge, and $Q=T_{3}+Y$ is the electric
charge.}%
\label{Table2}%
\begin{ruledtabular}
\begin{tabular}{ccccccc}
& $\phi^+$ & $\phi^0$ & $\eta$ & $\chi^{++}$ & $\chi^+$ & $\chi^0$ \\
\hline
$T_3$  & ${1 \over 2}$ & $-{1 \over 2}$ & 0 & 1 & 0 &$-1$  \\
$Y$ & ${1 \over 2}$ & ${1 \over 2}$ & 0 & 1 & 1 & 1    \\
$X$ &$ (0,2)$ & $(0,2)$ & 2 & $(0,-2)$ & $(0,-2)$ & $(0,-2)$   \\
\end{tabular}
\end{ruledtabular}\end{table}

The Yukawa potential with distinct charges takes the form.%

\begin{equation}
y_{1}\mathrm{{Tr}}\left(  \overline{E_{L}^{c}}E_{L}\right)  \eta_{(2)}%
+y_{2}\overline{L_{L}}\phi_{(2)}\ell_{R}+y_{3}\overline{L_{L}^{c}}i\sigma
_{2}\chi_{(-2)}L_{L}+y_{4}\overline{L_{L}^{c}}i\sigma_{2}E_{L}\phi_{(0)}%
+y_{5}\mathrm{{Tr}}\left(  \overline{E_{L}}\chi_{(0)}\right)  \ell_{R}+h.c.,
\end{equation}
where the indices in the lower brackets of the Higgses denote $U(1)_{X}$
charges of the Higgses.

A discrete $Z_{2}$ symmetry could be imposed to explain a certain
phenomenology, dark matter. If $E$ is odd and all other particles are even
under $Z_{2}$ symmetry, this would prevent the exotic leptons from coupling
with the ordinary leptons and the neutral lepton $E^{0}$ becomes stable, and
thus could be a dark matter candidate. The Yukawa potential with $Z_{2}$
symmetry is given by%

\begin{equation}
y_{1}\mathrm{{Tr}}\left(  \overline{E_{L}^{c}}E_{L}\right)  \eta_{(2)}%
+y_{2}\overline{L_{L}}\phi_{(2)}\ell_{R}+y_{3}\overline{L_{L}^{c}}i\sigma
_{2}\chi_{(-2)}L_{L}+h.c..
\end{equation}

The Yukawa potential for the quark sector may be built in an analogous way.%

\begin{equation}
y_{6}\overline{Q_{L}}\widetilde{\phi_{(0)}}u_{R}+y_{7}\overline{Q_{L}}%
\phi_{(0)}d_{R}+h.c.,
\end{equation}
where $Q=\left(  u,d\right)  ^{T}$ is the quark doublet and $\widetilde
{\phi_{(0)}}=i\sigma_{2}\phi_{(0)}^{\ast}$. The hypercharge combinations in
the potential prohibits the couplings between quarks and Higgs triplets. Since
quarks receive masses only from $\phi_{(0)}$, there is no tree level
flavor-changing neutral currents. Note that leptons and quarks interact with
two distinct Higgs doublets, which is different from the standard two Higgs
doublet model (2HDM) where one Higgs couples to the up-type quarks and the
other couples to the charged leptons and down-type quarks.

The size of the couplings and vacuum expectation values (VEVs) may be
approximately constrained with the known experimental measurements. The VEV of
$\phi_{(0)}$, $\left\langle \phi_{(0)}\right\rangle $, must be of the order of
$100$ GeV to meet the top quark mass, and $\ \left\langle \phi_{(2)}%
\right\rangle $ must be $1\sim100$ GeV to satisfy the $\tau$-lepton mass and
the known SM VEV\footnote{$\left\langle \phi\right\rangle =\sqrt{\left\langle
\phi_{(0)}\right\rangle ^{2}+\left\langle \phi_{(2)}\right\rangle ^{2}}%
\simeq174$ GeV$.$}. The Higgs triplet VEV $\left\langle \chi\right\rangle $
must be very small compared to the Higgs doublet VEV, since the $\rho$
parameter predicted by the SM is consistent with the experimental measurement
in high precision \cite{Cer07}. The neutrino mass may be generated at the tree
level in this model. The mass matrix of neutral leptons is%

\[
\mathcal{M}_{\nu E}=%
\begin{pmatrix}
m_{\nu} & y_{4}\left\langle \phi_{(0)}\right\rangle \\
y_{4}\left\langle \phi_{(0)}\right\rangle  & M_{E}%
\end{pmatrix}
,
\]
where $m_{\nu}\equiv y_{3}\left\langle \chi_{(-2)}\right\rangle $ and
$M_{E}\equiv y_{1}\left\langle \eta\right\rangle $. The nature of the neutrino
is not known; however, we have approximately predicted the size of the
neutrino mass. We take the exotic lepton $E$ at the weak scale, so the
coupling $y_{4}$ must be very small\footnote{According to the famous canonical
seesaw mechanism, the order unity coupling is assumed, with the scale of new
physics of 10$^{13}$GeV. However, we relax the constraint, as the coupling
$y_{4}$ could approximately be of the order of the electron Yukawa coupling
($\sim10^{-6}$).} since $\left\langle \phi_{(0)}\right\rangle $ is of the
order of $100$ GeV. The seesawlike mechanism is applicable to generate the
neutrino mass. If $Z_{2}$ symmetry is imposed, $y_{4}=0$. The neutrino mass
may be taken as $m_{\nu}$, where $y_{3}$ and/or $\left\langle \chi
_{(-2)}\right\rangle $ be sized for the neutrino mass. For either case, we
predict Majorana-type neutrinos in this model. Since $\left\langle
\chi\right\rangle $ and the coupling $y_{4}$ are small, we can consider that
the massive leptons are in the mass eigenstates for the Yukawa potential of (3).

The Higgs potential is also amenable to the gauge invariance with the extra
$U(1)$.%

\begin{align}
V  &  \supset V_{2HDM}+\left\{  \mu_{1}\phi_{(0)}^{T}\chi_{(0)}^{\dag}%
\phi_{(0)}+\mu_{2}\phi_{(2)}^{\dag}\chi_{(-2)}^{\dag}\phi_{(0)}+h.c.\right\}
\nonumber\\
&  +\left\{  \lambda_{1}\phi_{(2)}^{\dag}\phi_{(0)}\mathrm{{Tr}}\left(
\chi_{(-2)}^{\dag}\chi_{(0)}\right)  +\lambda_{2}\phi_{(2)}^{\dag}\sigma
^{a}\phi_{(0)}\mathrm{{Tr}}\left(  \chi_{(-2)}^{\dag}\sigma^{a}\chi
_{(0)}\right)  +h.c.\right\}  ,
\end{align}
where $V_{2HDM}$ stands for the Higgs potential involving only Higgs doublets,
and the functional form is the same as the 2HDM with $Z_{2}$ symmetry. In
addition to the two complex trilinear couplings, the two complex quartic
couplings are possible, those involving the CP violation phenomenology. The
phenomenology with two complex trilinear coupings can be found in
Ref.\cite{Ema98}\footnote{They assigned the Higgs triplets of the order of
$10^{13}$GeV to explain neutrino masses and Baryogenesis via Leptogenesis.
However the Higgs triplets are assumed to have masses of the order of weak
scale to explain the interesting phenomenology in our scenario.}, and the
complex quartic couplings are related to the electric dipole moment of
fermions which will be discussed as a part of this letter. The other
interaction terms are trivial and almost irrelevant to the phenomenology.

\section{$Z^{\prime}$ discovery limit}

The interactions of the $Z^{\prime}$ boson with the fermions are described by%

\begin{equation}%
{\displaystyle\sum\limits_{f}}
z_{f}^{\prime}g_{Z^{\prime}}Z_{\mu}^{\prime}\overline{f}\gamma^{\mu}f,
\end{equation}
where $f=E_{L},Q_{L},L_{L},u_{R},d_{R},e_{R}$ are the lepton and quark fields
and $z_{f}^{\prime}$ is the gauge charge corresponding to the fermion.

The leptonic decays $Z^{\prime}\rightarrow\ell^{+}\ell^{-}(e^{+}e^{-}$ and
$\mu^{+}\mu^{-})$ provide the most distinctive signature for observing the
$Z^{\prime}$ signal at the hadron colliders. The cross section of the
$p{\bar{p}}$ collision in the $\ell^{+}\ell^{-}$ channel can be calculated at
the narrow width $Z^{\prime}$ pole in the center-of-momentum (CM) frame. The
hadronic cross section is given by%

\begin{equation}
\sigma(Z^{\prime})=K\sum_{q,{\bar{q}}}\int_{0}^{1}dx_{1}dx_{2}(f_{q}^{p}%
(x_{1})f_{\overline{q}}^{\overline{p}}{(x_{2})}+f{_{\overline{q}}^{p}(x_{1}%
)}f_{q}^{\overline{p}}(x_{2})){\hat{\sigma}(Z^{\prime}),}%
\end{equation}
where ${\hat{s}=x}_{1}x_{2}s$ is the partonic fraction of $s$, $f(x)$'s are
the partonic distribution functions (PDFs) and the sum is performed over all
the light quarks. $K$ is the QCD correction factor ($\sim$ 1.3) \cite{Mcar04},
which accounts for higher order QCD corrections. The partonic cross secion
${\hat{\sigma}(Z^{\prime})}$ is calculated in a sum over the spins of the
final states and an average over the spins and colors of the initial states.%

\begin{equation}
{\hat{\sigma}}(Z^{\prime})={\frac{\pi z_{f}^{\prime2}g_{Z^{\prime}}^{2}}{48}%
}\delta({\hat{s}}-M_{Z^{\prime}}^{2}).
\end{equation}

Eq.(8) and (9) lead to the hadronic cross section in the $\ell^{+}\ell^{-}$ channel.%

\begin{equation}
\sigma(Z^{\prime})\cdot Br_{\ell^{+}\ell^{-}}=K{\frac{\pi z_{f}^{\prime
2}g_{Z^{\prime}}^{2}}{48s}}\sum_{q,{\bar{q}}}\int_{\frac{m_{Z^{\prime}}^{2}%
}{s}}^{1}{\frac{dx}{x}}\left(  f_{q}^{p}(x)f_{\overline{q}}^{\overline{p}%
}\left(  {\frac{M_{Z^{\prime}}^{2}}{xs}}\right)  +f{_{\overline{q}}^{p}%
(x)}f_{q}^{\overline{p}}\left(  {\frac{M_{Z^{\prime}}^{2}}{xs}}\right)
\right)  \cdot Br_{\ell^{+}\ell^{-}},
\end{equation}
where $Br_{\ell^{+}\ell^{-}}$ is the branching ratio of $Z^{\prime}$ to
$\ell^{+}\ell^{-}$. We may take $z_{f}^{\prime2}\simeq1,$ since precision
measurements of $Z$-pole observables predict the small $Z-Z^{\prime}$ mixing
angle$(\leq10^{-3})$ \cite{Cer07}.%

\begin{figure}
[ptb]
\begin{center}
\includegraphics[
trim=0.000000in 0.000000in 0.000000in -0.247569in,
height=9.5355cm,
width=13.0479cm
]%
{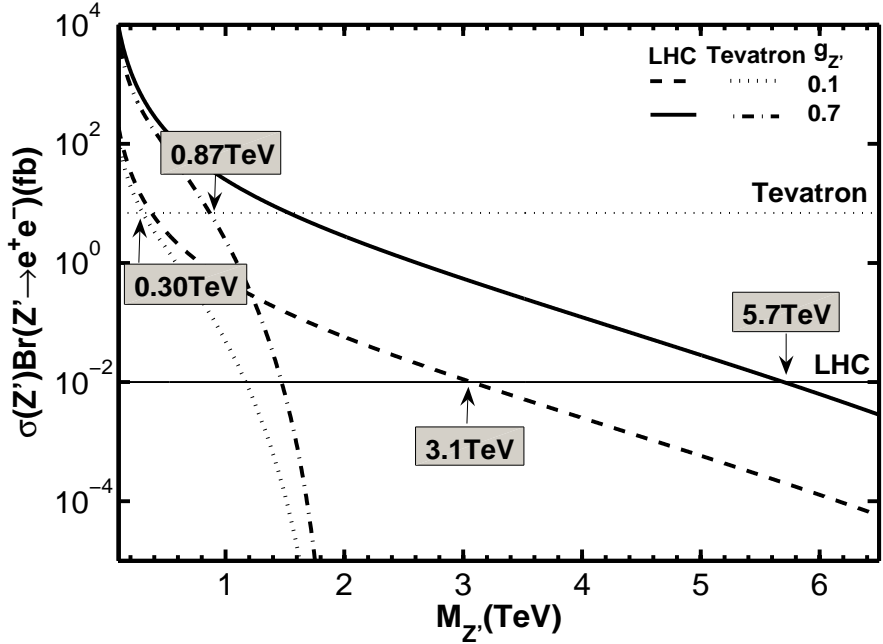}%
\caption{The $Z^{\prime}$ discovery limit at the Tevatron ($\sqrt{s}=1.96$ TeV
and $L=1.3$ $fb^{-1}$) and the LHC($\sqrt{s}=14$ TeV and $L=100$ $fb^{-1}$).
The horizontal lines indicate the experimental sensitivities, and the bold
lines are predictions of the cross section. The predictions are for the
coupling, $g_{Z^{\prime}}=0.1$ and $0.7$ (SM coupling). MRST LO PDFs
\cite{Mar02} are used. The intersections of the curves determine the lower
mass limits.}%
\label{fig1}%
\end{center}
\end{figure}

For $pp$ collision at the LHC, the proton PDF takes the place of the
antiproton PDF. Fig.1 shows the predicted cross sections with the present
experimental sensitivity at the Tevatron Run II\footnote{The CDF Collaboration
\cite{CDF07} has set the better luminosity for $\sigma(Z^{\prime})\cdot
Br_{\ell^{+}\ell^{-}}$ than the D\O \ \cite{D007} in some reason, so it is
considered for the CDF collider.} and the projected experimental sensitvity at
the LHC \cite{LHC05}. The actual experimental analysis shows an experimental
line with a more complicated structure than the horizontal line in the figure.
For a nonzero background\footnote{The non-zero background is roughly taken
from Ref.\cite{CDF07}, which all the expected backgrounds are considered. The
most significant source of background in this channel is the SM Drell-Yan
process via $Z/\gamma^{\ast}$ as reported in Ref. \cite{CDF07}.},
$N_{Z^{\prime}}=3$ events are excluded at the Tevatron. The $Z^{\prime}$
discovery limits are $300$ GeV, $870$ GeV for $g_{Z^{\prime}}=0.1,0.7$ at the
Tevatron, and the LHC may probe $Z^{\prime}$ upto $3.1$ TeV, $5.7$ TeV for
$g_{Z^{\prime}}=0.1,0.7$. Since the $U(1)_{X}$ gauge charge of the Higgs
singlet $\eta$ is $2$, $M_{Z^{\prime}}\simeq2g_{Z^{\prime}}\left\langle
\eta\right\rangle $. We predict the lower limit of the extra $U(1)$ symmetry
breaking to be around $200\sim800$ GeV at the Tevatron (CDF detector).

\section{muon anomalous magnetic moment}

The deviation of the current experimental value from the SM prediction is
approximately $3.0\sigma$ and the numerical deviation is $\Delta a_{\mu
}=27.5(8.4)\times10^{-10}$ \cite{Mda07} or $27.7(9.3)\times10^{-10}$
\cite{Fdo08}. The experimental value is the measurement of the BNL experiment
\cite{Muo06}. We investigate one- and two-loop contributions.%

\begin{figure}
[ptb]
\begin{center}
\includegraphics[
trim=0.000000in 0.000000in -0.412326in 0.000000in,
height=1.5463in,
width=3.614in
]%
{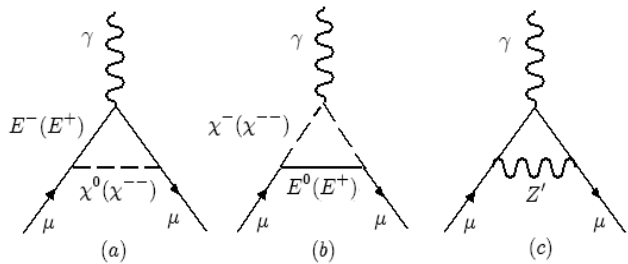}%
\caption{The one-loop contributions to $a_{\mu}$ involving the extra
particles, $E,\chi,$ and $Z^{\prime}$. }%
\label{fig2}%
\end{center}
\end{figure}

The diagrams of Fig.2 display one-loop contributions involving the new
particles, $E$, $\chi,$ and $Z^{\prime}$. The relevant interaction Lagrangian
for diagrams (a) and (b) of Fig. 2 comes from the $y_{5}$-term of the Yukawa
potential of (3). The states $\chi_{(0)},\chi_{(-2)}$ may be rotated into the
mass eigenstates $\chi_{\ell},\chi_{h}$, where $\chi_{\ell}$ and $\chi_{h}$
are the light and heavy mass eigenstates. The rotational angle is determined
in the Higgs potential. However, the couplings with the Higgs triplet are free
parameters, so we redefine the new couplings in the mass eigenstates. The
relevant Lagrangian for the light scalar state $\chi_{\ell}$ is given by\ %

\begin{equation}
-y\left(  \overline{E_{L}^{0}}\chi_{\ell}^{+}\mu_{R}+\overline{E_{L}^{-}}%
\chi_{\ell}^{0}\mu_{R}-\overline{E_{L}^{+}}\chi_{\ell}^{++}\mu_{R}%
+h.c.\right)  ,
\end{equation}
where $y$ is the Yukawa coupling in the mass eigenstates of $\chi$. The
Lagrangian for the heavy mass eigenstates can be given in the same fashion.
The $y_{4}$-term with which the Higgs doublet is involved is neglected due to
the small coupling constrained by the neutrino mass.

The contribution of Fig.2(a) is negligible, since the particles ($E^{+}%
$,$E^{-}$) on the line which are hooked up by the photon have opposite
electric charges. We calculated the contribution of Fig.2(b), and it is given by%

\begin{equation}
\Delta a_{\mu}^{(\text{one})}=\frac{3y^{2}}{8\pi^{2}}\left(  \frac{m_{\mu}%
}{M_{E}}\right)  f\left(  \frac{M_{\chi}^{2}}{M_{E}^{2}}\right)
\simeq4.03\times10^{-6}\cdot y^{2}\left(  \frac{1\text{TeV}}{M_{E}}\right)
f\left(  \frac{M_{\chi}^{2}}{M_{E}^{2}}\right)  ,
\end{equation}
where the prefactor of 3 comes from the electric charges of $\chi^{\pm}%
,\chi^{\pm\pm}$.

The corresponding one-loop function is%

\begin{equation}
f(z)=\int_{0}^{1}dx\frac{(1-x)x}{zx+1-x}=\frac{1-z^{2}+2z\ln z}{2(1-z)^{3}}%
\end{equation}
which has asymptotic behaviors,%

\begin{equation}
f(z)\longrightarrow\left\{
\begin{array}
[c]{cc}%
\frac{1}{6} & \text{as }z=1\ ,\\
{\frac{1}{2z}-}\frac{\ln z}{z^{2}} & \text{ for }z\gg1\ ,\\
\frac{1}{2}+z\ln z & \text{for }z\ll1\ .
\end{array}
\right.  .
\end{equation}

We neglect the contribution from the other scalars (called the heavy scalars),
since those scalars are split into light and heavy mass eigenstates, in
general, and the one-loop function behaves $f(z)\rightarrow0$ as
$z\rightarrow\infty$. \ Furthermore, the large splitting is necessary to
generate the sizable electric dipole moment, that will be discussed in the
next section. The $M_{\chi}$ or $M_{\chi_{\ell}}$ implies the mass of the
light scalar in this letter.%

\begin{figure}
[ptb]
\begin{center}
\includegraphics[
trim=0.000000in 0.000000in 0.012447in 0.000000in,
height=9.8189cm,
width=14.0518cm
]%
{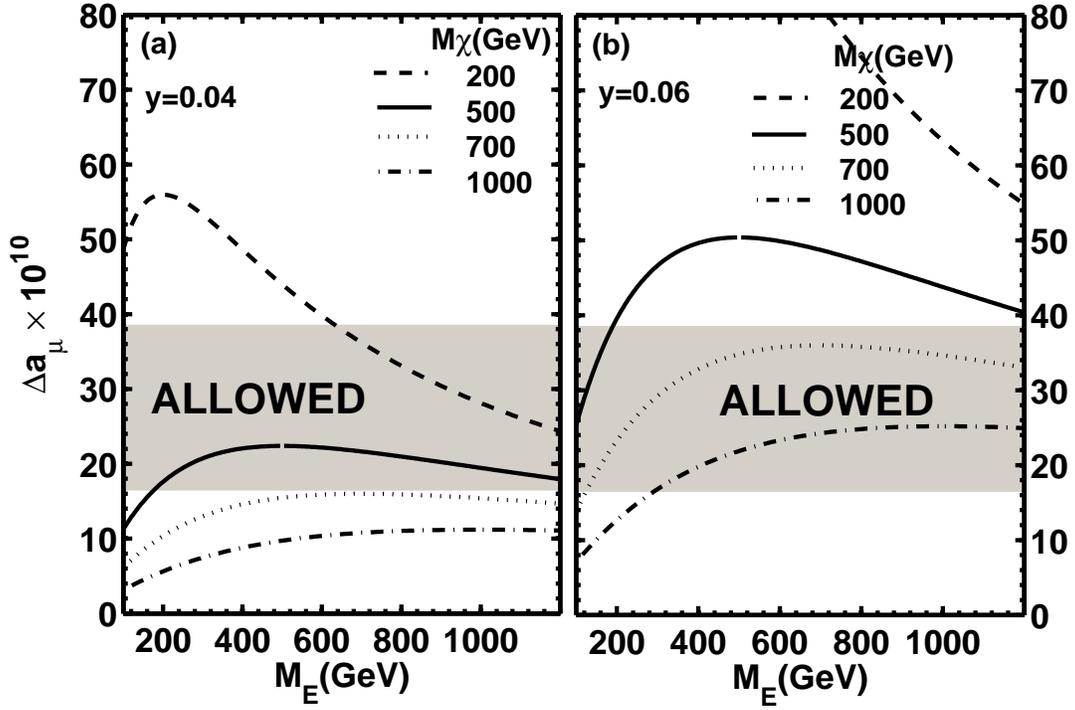}%
\caption{$\Delta a_{\mu}$ as a function of the exotic lepton mass $M_{E}$ for
various values of $M_{\chi}$ at the one-loop level. }%
\label{fig3}%
\end{center}
\end{figure}

Fig.3 shows the predictions of the anomalous magnetic moment for
$0.1$TeV$<M_{E},M_{\chi}<1$TeV. The range of deviations from the SM is
presented in the dark "allowed" band \cite{Mda07}. The predictions are in the
allowed band around the Yukawa coupling $y=0.05$. Since $\Delta a_{\mu
}^{(\text{one})}\sim y^{2}/M_{E},$ the Yukawa coupling $y$ is very sensitive
to the deviation $\Delta a_{\mu}$. Besides the above region, a possible
scenario is $M_{E}\approx M_{\chi}>1$TeV for the Yukawa coupling $y>0.06$. The
contribution by the $Z^{\prime}$ gauge boson of Fig.2(c) is negligible, since
$\Delta a_{\mu}\sim m_{\mu}^{2}/M_{Z^{\prime}}^{2}$.%

\begin{figure}
[ptb]
\begin{center}
\includegraphics[
trim=0.000000in 0.000000in 0.000000in -0.071326in,
height=1.9571in,
width=3.7542in
]%
{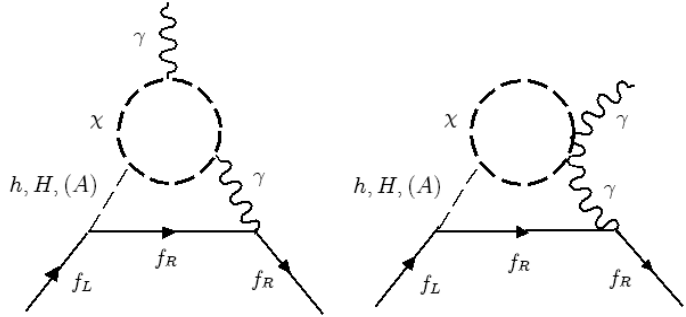}%
\caption{Two-loop contributions to $a_{\mu}$ $(d_{e})$ (mirror graphs are not
displayed.).}%
\label{fig4}%
\end{center}
\end{figure}

If $Z_{2}$ symmetry is imposed, there is no one-loop contribution to explain
the deviation. We consider the two-loop contribution via the Barr-Zee type of
mechanism, which is depicted in Fig.4. The relevant Lagrangian to induce the
Barr-Zee two-loop contribution is given by%

\[
-\frac{\sqrt{2}m_{\mu}r_{\mathcal{H}}}{v}\overline{\mu}\mathcal{H}\mu
-\frac{\lambda_{+}v}{\sqrt{2}}\mathcal{H}\left(  \chi_{\ell}\chi_{\ell}%
+\chi_{h}\chi_{h}\right)  ,
\]
where $\mathcal{H}=h$ or $H$, $v=\sqrt{2}\left\langle \phi\right\rangle ,$ and
$\lambda_{+}$ is the coupling in the mass eigenstates of $\chi$ for
$\mathcal{H}$. The rotational angles\footnote{Coventionally, the rotational
angle between the neutral Higgses in 2HDM is denoted by the symbol $\alpha$.
But in this letter, the symbol $\alpha$ is used for the electric fine
structure constant, so we use the symbol $\beta_{h}$ for the rotational
angle.} $r_{h}=-\sin\beta_{h}/\cos\beta$ and $r_{H}=\cos\beta_{h}/\cos\beta$
to the muon are the same as in the standard 2HDM, since the scalar $\phi
_{(2)}$, which is consistent with the scalar to couple to the charged leptons
in the 2HDM, couples to the muon. There is no contribution from the CP-odd
Higgs (pseudoscalar) $A$ because the interaction with the CP-odd Higgs
violates CP symmetry, so the effect of the CP-odd Higgs involves the electric
dipole moment.

The contribution of two loops is given by%

\begin{align}
\Delta a_{\mu}^{(\text{two})}  &  \simeq-\sum_{\mathcal{H},\chi}\frac{\alpha
m_{\mu}^{2}}{16\pi^{3}}\frac{Q_{\chi}^{2}r_{\mathcal{H}}\lambda_{+}%
}{m_{\mathcal{H}}^{2}}\left[  F\left(  \frac{M_{\chi_{\ell}}^{2}%
}{m_{\mathcal{H}}^{2}}\right)  +F\left(  \frac{M_{\chi_{h}}^{2}}%
{m_{\mathcal{H}}^{2}}\right)  \right] \nonumber\\
&  =-2.07\times10^{-11}\cdot\sum_{\mathcal{H}=h,H}\lambda r_{\mathcal{H}%
}\left(  \frac{200\text{GeV}}{m_{\mathcal{H}}}\right)  ^{2}\left[  F\left(
\frac{M_{\chi_{\ell}}^{2}}{m_{\mathcal{H}}^{2}}\right)  +F\left(
\frac{M_{\chi_{h}}^{2}}{m_{\mathcal{H}}^{2}}\right)  \right]  .
\end{align}

Note that $\sum Q_{\chi}^{2}=5$ due to singly and doubly charged scalars in
the inner loop. The two-loop function is%

\begin{equation}
F\left(  z\right)  =\int_{0}^{1}dx\frac{x(1-x)}{z-x(1-x)}\ln\left[
\frac{x(1-x)}{z}\right]
\end{equation}
which has asymptotic behaviors,%

\begin{equation}
F(z)\longrightarrow\left\{
\begin{array}
[c]{cc}%
-{0.344} & \text{as }z=1\ ,\\
-{\frac{1}{6z}}\ln z-{\frac{5}{18z}} & \text{ for }z\gg1\ ,\\
(2+\ln z) & \text{for }z\ll1\ .
\end{array}
\right.  \ .
\end{equation}

The Barr-Zee two-loop contributions, according to Eq.(15), are suppressed by
the muon mass and the loop factor, and thus the large $r_{\mathcal{H}}$ and
the small $m_{\mathcal{H}}$ are necessary. The lower limit of the light Higgs
boson mass is around $44$ GeV for $r_{h}\simeq\tan\beta$ from the LEP
\cite{Opal01}, but the light Higgs boson keeps the same lower limit of the SM
Higgs boson, $113.5$ GeV, for $r_{H}\simeq\tan\beta.$ The case for
$r_{h}\simeq\tan\beta$ is taken. We can approach these analyses in the 2HDM
since the VEVs of the Higgs triplets have the small size. Besides, the doubly
charged scalar $\chi^{++}$ in the inner loop gives the main contribution to
the deviaton due to its double electric charge. The lower limit of the doubly
charged scalar, around $120$ GeV from the Tevatron \cite{CDF04} and the LEP
\cite{L303}, is considered. \ %

\begin{figure}
[ptb]
\begin{center}
\includegraphics[
trim=0.000000in 0.000000in 0.000000in -0.158308in,
height=10.3395cm,
width=14.3527cm
]%
{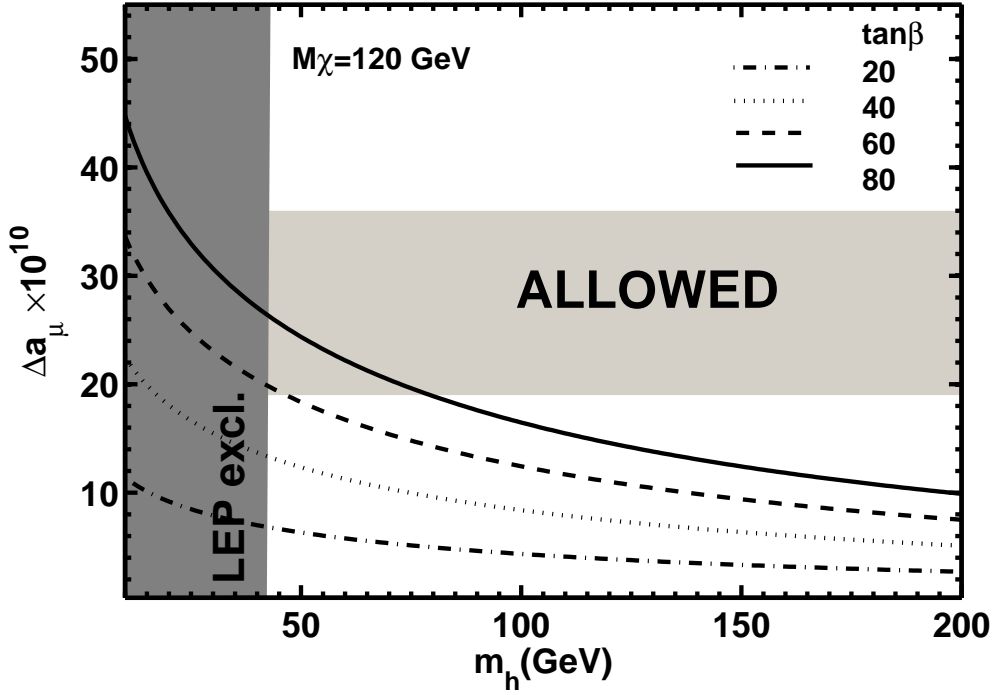}%
\caption{$\Delta a_{\mu}$ as a function of the light Higgs boson mass $m_{h}$
at the two-loop level for various values of $\tan\beta$.}%
\label{fig5}%
\end{center}
\end{figure}

Fig.5 shows the predictions for the Barr-Zee two-loop contribution, $\Delta
a_{\mu}^{(\text{two})}$, as a function of the light Higgs boson mass $m_{h}$.
To predict the two-loop contribution $\Delta a_{\mu}^{(\text{two})}$, we
assume a coupling $\lambda_{+}$ of the same size as the SM Higgs quartic
coupling for the SM Higgs of 120 GeV. The predictions barely reside in the
allowed region.

\section{Electron Electric Dipole Moment}

The EDM of fermions predicted by the standard model is extremely small
compared to the present experimental bounds. Another mechanism beyond the SM
has been required to induce the sizable EDM. There are also explicit CP
violation interactions related to the Barr-Zee two-loop mechanism
\cite{Smba90,Dar99,Jheo08} for the EDM in this model. Since the interaction
must involve the CP violation, it is comprised of only the CP-odd Higgs
(pseudoscalar) $A$. The irreducible CP phase appears in the
diagonalization\footnote{The detailed process for diagonalization of mass
matrix by the unitary transformation can be found in Ref.\cite{Jheo08}.} of
the mass matrix for the Higgs triplets in the Higgs potential of (6). If we
introduce the new phenomenological couplings, the relevant interaction
Lagrangian is given by%

\[
\frac{\sqrt{2}m_{\mu}r_{A}}{v}\overline{e}i\gamma^{5}Ae-\frac{\lambda_{-}%
v}{\sqrt{2}}A\left(  \chi_{\ell}\chi_{\ell}-\chi_{h}\chi_{h}\right)  ,
\]
where $r_{A}=\tan\beta$ is the rotational angle, and $\lambda_{-}=\lambda
\sin\delta$ where $\sin\delta$ is the CP-violation effect which comes from
combinations of the complex quartic couplings in the potential of (6). The
Barr-Zee diagrams were well calculated in many papers to induce the sizable
electric dipole moment, and the result is identical to the Barr-Zee two-loop
contribution of the anomalous magnetic moment, except for CP-violation effect.
The electron electric dipole moment results in%

\begin{align}
\left(  \frac{d_{e}}{e}\right)  ^{\gamma}  &  =-\sum_{\chi}\frac{\alpha m_{e}%
}{32\pi^{3}}\frac{Q_{\chi}^{2}r_{A}\lambda_{-}}{m_{A}^{2}}\left[  F\left(
\frac{M_{\chi_{\ell}}^{2}}{m_{A}^{2}}\right)  -F\left(  \frac{M_{\chi_{h}}%
^{2}}{m_{A}^{2}}\right)  \right] \nonumber\\
&  =-9.25\times10^{-27}\cdot\left(  \frac{200\text{GeV}}{m_{A}}\right)
^{2}\lambda\sin\delta\tan\beta\left[  F\left(  \frac{M_{\chi_{\ell}}^{2}%
}{m_{A}^{2}}\right)  -F\left(  \frac{M_{\chi_{h}}^{2}}{m_{A}^{2}}\right)
\right]  \text{ \ }(cm),
\end{align}
where the two-loop function is given in Eq.(16) ; also note that $\sum
Q_{\chi}^{2}=5$ due to singly and doubly charged scalars from the Higgs
triplets. The electron EDM results in the difference between two contributions
from the light and heavy scalars, $\chi_{\ell}$ and $\chi_{h}$. The
contribution from the heavy scalar is neglected, since the two-loop function
behaves like $F\left(  z\right)  \rightarrow0$ as $z\rightarrow\infty$.%

\begin{figure}
[ptb]
\begin{center}
\includegraphics[
trim=0.000000in 0.000000in -0.190854in 0.000000in,
height=3.7542in,
width=5.5305in
]%
{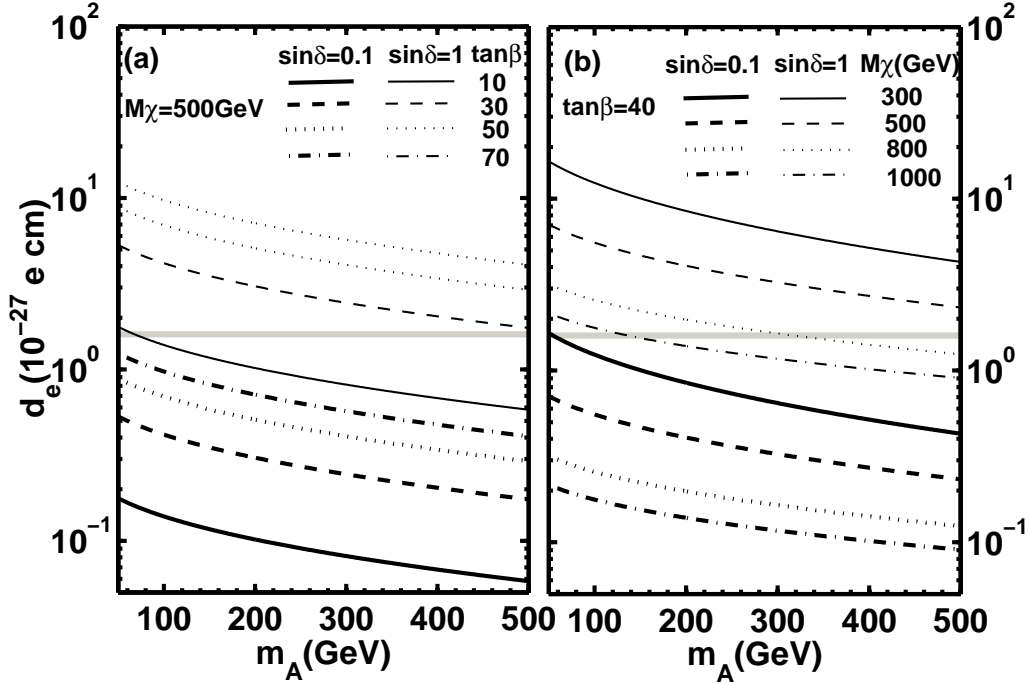}%
\caption{Numerical estimates of the EDMs as a function of the CP-odd (or
pseudoscalar) Higgs boson mass for various values of $\tan\beta$ and $M_{\chi
}$. Also shown the predictions for CP-phase, $0.1\leq\sin\delta\leq1$
$(6^{\circ}\leq\arg(A)\leq90^{\circ})$. The horizonal line indicates the
current 90\% C.L. experimental bound \cite{Bcr02}.}%
\label{fig6}%
\end{center}
\end{figure}

In order to predict the electron electric dipole moment numerically, we also
assume the coupling $\lambda$ of the same size as the SM Higgs quartic
coupling for the SM Higgs of 120 GeV. Fig.6 shows the predictions of the
electron electric dipole moment as a function of the CP-odd Higgs (or
pseudoscalar) mass with the current 90\% C.L. experimental bound \cite{Bcr02}.
The sizable contributions are expected for the moderate size of the CP phase,
$0.1\leq\sin\delta\leq1$ $(6^{\circ}\leq\arg(A)\leq90^{\circ})$.

\section{Conclusions}

The model (Lagrangian) with a peculiar extra $U(1)$, that Barr and Dosner
suggested, has clearly been presented. The gauge charges of the extra $U(1)$
give a strong constraint to build the Lagrangians. $Z^{\prime}$ discovery
limits are estimated and predicted at the Tevatron and the LHC. The discovery
limit at the Tevatron (CDF detector) gives the lower limit of the extra $U(1)$
symmetry breaking scale, approximately $200\sim800$ GeV. The muon anomalous
magnetic moment could be explained at the one-loop level for a Yukawa coupling
around $0.05$. If we allow masses of the new particles to be more than 1 TeV,
the larger Yukawa coupling is possible. However, smaller Yukawa couplings are
prohibited by the discovery limits of new particles at the Tevatron and the
LEP. The muon anomalous magnetic moment could also be explained at the
two-loop level, but the region of parameters is very narrow. There are
explicit CP-violation interactions in this model. A sizable electron electric
dipole moment is expected for a moderately sized CP phase, $0.1\leq\sin
\delta\leq1$, $(6^{\circ}\leq\arg(A)\leq90^{\circ})$ via the Barr-Zee mechanism.

\end{document}